\documentstyle[12pt,epsfig]{article}
  \def\thebibliography#1{\center{\bf REFERENCES}\list
   {[\arabic{enumi}]}{\settowidth\labelwidth{[#1]}\leftmargin\labelwidth
   \advance\leftmargin\labelsep
   \usecounter{enumi}}
   \def\newblock{\hskip .11em plus .33em minus -.07em}
   \sloppy
   \sfcode`\.=1000\relax}
  
  \newcounter{cap}
  {\begin{list}{Figure \arabic{cap}\hfil}{\usecounter{cap}
  \settowidth{\labelwidth}{Figure #1}%
  \setlength{\leftmargin}{\labelwidth}%
  \addtolength{\leftmargin}{\labelsep}%
  \setlength{\parsep}{2mm plus 1mm minus 1mm}
  \setlength{\itemsep}{3mm plus 2mm minus 2mm}
  }}%
  {\end{list}}
  {\begin{list}{}{\settowidth{\labelwidth}{#1}%
  \setlength{\leftmargin}{\labelwidth}%
  \addtolength{\leftmargin}{\labelsep}%
  \setlength{\itemsep}{0pt plus 1pt}
  \setlength{\parsep}{0pt plus 1pt}
  \setlength{\topsep}{0pt plus 1pt}
  \setlength{\partopsep}{0pt plus 1pt}
  \setlength{\parskip}{2mm plus 1mm minus 1mm}
  }}%
  {\end{list}}
\begin{document}
\parindent 1.3cm
\thispagestyle{empty}   
\vspace*{-3cm}
\noindent
\vspace{1.5cm}

\def\arccot{\mathop{\rm arccot}\nolimits}
\def\sd{\strut\displaystyle}

\begin{flushright}
NORDITA-96/8 N/P \\
UAB-FT-96/382
\end{flushright}

\vspace{1.5cm}

\begin{center}
\begin{bf}
\noindent
CHIRAL PERTURBATION THEORY PREDICTIONS FOR 
$\eta \to \pi^+ \pi^- \pi^0 \gamma$
\end{bf}
  \vspace{2cm}\\
A. BRAMON$^{a)}$, P. GOSDZINSKY$^{a,b,c)}$ and S. TORTOSA$^{a)}$
   \vspace{0.1cm}\\
$^{a)}$ \ Grup de F\'\i sica Te\`orica, Universitat Aut\`onoma de Barcelona,\\ 
08193 Bellaterra (Barcelona), Spain\\
$^{b)}$ \ Nordita, Blegdamsvej 17, DK-2100 Copenhagen \O , Denmark\\
$^{c)}$ \ IFAE, Universitat Aut\`onoma de Barcelona,\\ 
08193 Bellaterra (Barcelona), Spain\\
   \vspace{5.0cm}

{\bf ABSTRACT}
\end{center}
The $\eta \to \pi^+ \pi^- \pi^0 \gamma$ decay is 
discussed in the general context of Chiral Perturbation Theory (ChPT),
assuming that the low--energy constants 
(counter--terms) are saturated by vector-meson resonances.
The $\eta \to \pi^+ \pi^- \pi^0 \gamma$ amplitude can be 
separated in two distinct pieces: the inner bremsstrahlung,
$A^{(IB)}$, and the structure dependent (or direct emission), 
$A^{(SD)}$, amplitudes. 
The former -- which essentially contains the same physics as
$A(\eta \to \pi^+ \pi^- \pi^0)$ -- is found to dominate over 
the second one -- which looks more interesting from 
the ChPT point of view.

\newpage

{\bf Introduction}

At low energies, Chiral Perturbation Theory (ChPT) 
provides an accurate description of the 
strong and electroweak interactions of pseudoscalar mesons, $P$,
in terms of a perturbative series expansion \cite{GASSER}. 
At lowest order, ChPT is essentially equivalent to Current Algebra 
(CA) \cite{DASHEN} but, at higher orders, loop effects appear 
restoring unitarity and giving rise to both 
finite corrections and divergences. The former are known to 
improve the lowest order (or CA) predictions, while the latter 
require the introduction of counterterms and 
destroy the renormalizability of the theory. The finite part of these 
counterterms, in turn, can be related to other successful aspects of 
hadron physics, such as the classical notion of Vector Meson Dominance 
(VMD). In this note we study the $\eta \to \pi^+ \pi^- \pi^0 \gamma$ 
decay in the context of ChPT with VM dominated counterterms.
\par
The amplitude for the $\eta \to \pi^+ \pi^- \pi^0 \gamma$ decay 
contains two distinct pieces receiving two distinct types of contributions,
$A( \eta \to \pi^+ \pi^- \pi^0 \gamma) = A^{(IB)} + A^{(SD)}$. 
The first piece -- the inner bremsstrahlung amplitude $ A^{(IB)} $
-- proceeds through the hadronic and 
G-parity violating $\eta \to \pi^+ \pi^- \pi^0$ decay, accompanied 
by the emission of an isovector (positive G-parity) photon. This part 
of the amplitude presents an infrared divergence and it is obviously 
dominated by soft photon emission. It can be deduced from 
Low's theorem and, therefore, $A^{(IB)}$ essentially contains the 
same physics as the purely hadronic $\eta \to \pi^+ \pi^- \pi^0$ 
amplitude carefully discussed by Gasser and Leutwyler \cite{GASSER}. 
The second piece -- the structure dependent (or direct emission) 
amplitude $ A^{(SD)}$ -- looks {\it a priori} more 
interesting from the ChPT point of view. It involves an isoscalar 
(negative G-parity) photon and proceeds without the suppression due to 
G-parity violation. Therefore there is no obvious reason to assume 
that this G-parity conserving part of the amplitude, $A^{(SD)}$,
has to be smaller than the first, G-parity violating one, 
$A^{(IB)}$, except for the low-energy end of the photonic spectrum.
\par
The same conclusion was reached long ago by Singer \cite{SINGER} 
in his VMD calculation, as well as in the more detailed CA analysis of 
ref. \cite{INTEMANN}, where $A^{(IB)}$ was predicted to be larger 
than $A^{(SD)}$ only for photon energies $E_{\gamma} \leq 15$ MeV, 
well below its maximum value $E_{\gamma}^{max} \simeq 120$ MeV.
These two and other old theoretical calculations for $A^{(SD)}$
(based on CA and/or VMD) \cite{SINGER,INTEMANN,old} lead to
\begin{equation}
\label{BR}
\frac {\Gamma ^{(SD)} (\eta \to \pi^+ \pi^- \pi^0 \gamma)}
{\Gamma (\eta \to \pi^0 \gamma \gamma)} \ = \ 0.23 \ \%, \ 0.24 \ \%, \ 
0.28 \ \%,
\end{equation}
although another similar (but oversimplified) analysis \cite{Baracca} 
lead to a negligible $A^{(SD)}$ and to smaller values for 
the ratio above. Combining the predictions (\ref{BR}) with present day 
data for the eta meson \cite{PDG} one obtains
\begin{equation}
\label{WIDTHBR}
\Gamma ^{(SD)}(\eta \to \pi^+ \pi^- \pi^0 \gamma) \ 
\simeq \ 2 \times 10^{-3}  \ eV, \ \ 
B. R.^{(SD)}(\eta \to \pi^+ \pi^- \pi^0 \gamma) \ \simeq  \ 1.6 \times 
10^{-6}.
\end{equation}
Up to now, this structure dependent part of the 
$\eta \to \pi^+ \pi^- \pi^0 \gamma$ decay has not been 
detected experimentally and only an upper limit  -- 
at 90 \% of confidence level -- is known \cite{THALER}.
\begin{equation}
\label{WIDTH}
\Gamma^{(SD)}(\eta \to \pi^+ \pi^- \pi^0 \gamma)\ \le \ 6 \times 10^{-4}
\ \Gamma (\eta \to all)\ \simeq 0.72 \ eV.  
\end{equation}
According to this discussion, the Saturne $\eta$-factory 
and the Daphne $\phi$-factory could in principle 
produce enough etas to allow 
for detection and analysis of the structure dependent part of the 
$\eta \to \pi^+ \pi^- \pi^0 \gamma$ amplitude.
This prompted us to perform the corresponding ChPT calculation 
following the lines of ref. \cite{pigg}, where the analog (and 
similarly complicated) $\eta \to \pi^0 \gamma \gamma$ decay was 
considered.

We will describe the nonet of pseudoscalar mesons $P$ in
terms of the $SU(3)$ octet and singlet matrices
\begin{equation}
\label{M_8}
P_8=\pmatrix{ 
{\sd\pi^0 \over \sd\sqrt 2}+{\sd\eta_8 \over \sd\sqrt{6}} &  \pi^+  &  K^+    \cr
\pi^-   & -{\sd\pi^0 \over \sd\sqrt 2} + {\sd\eta_8 \over \sd\sqrt{6}} &  K^0 \cr
 K^-    &     \bar K^0      & {-\sd 2\over \sd\sqrt{6} } \eta_8  \cr }, 
\qquad P_1={\sd{1\over \sqrt{3}}}\eta_1 \hbox{I,}
\end{equation}
which appear in the ChPT lagrangian through the parametrization
\begin{equation}
\label{SIGMA}
\Sigma\equiv\Sigma_8\Sigma_1=\Sigma_1\Sigma_8=\exp\left(\sd{2i\over 
f}(P_8+P_1)\right),
\end{equation}
with $f=132~\hbox{MeV}$ \cite{GASSER,PDG}. Following the lines of ref. 
\cite {pigg} and the reviews \cite{REV}, we will adopt a rather 
simplified treatment of the singlet component of the $\eta$ in ChPT. 
More precisely, we will assume that 
the physical $\eta$ particle participates from nonet symmetry with 
an $\eta - \eta'$ mixing given by
\begin{equation}
\label{MIXING}
\eta=\cos\theta~\eta_8 -\sin \theta~\eta_1 \simeq \sd{1\over\sqrt{3}}(u\bar
u +d \bar d - s \bar s), \quad \sin\theta\simeq -1/3,
\end{equation}
as deduced from conventional $\eta-\eta '$ phenomenology and 
from more recent treatments in the
ChPT context \cite{REV,BIJNENS1,BIJNENS2}.

The lowest order lagrangian of ChPT (order two in particle 
four--momenta or masses, $O(p^2)$) is  
\begin{equation}
\label{L2}
{L}_2=\sd{f^2\over 8} tr (D_\mu\Sigma D^\mu\Sigma^\dagger + \chi 
\Sigma^\dagger +\chi^\dagger \Sigma),
\end{equation}
apart from an extra singlet mass term for $\eta_1$ that should be 
added to account for the $U(1)_A$ problem, as discussed in \cite{REV}.
The first term in (\ref{L2}) contains the covariant derivative
$D_\mu \Sigma \equiv \partial_\mu \Sigma + i e A_\mu [Q,\Sigma],$
 with the photon field $A_\mu$ and the quark charge matrix $Q$  
[$Q=\hbox{diag}(2/3,-1/3,-1/3)$]. 
The non--derivative terms in Eq. 
(\ref{L2}), with $\chi=\chi^\dagger = B~\cal M$,
contain the quark mass matrix $\cal M$ 
[${\cal M}=\hbox{diag}(m_u,m_d,m_s)$] and lead to
\begin{equation}
\label{B}
\sd{{1\over 2}B={m_K^2\over m_u+m_s}= {m_\pi^2\over m_u+m_d}= {\Delta 
m_K^2\over m_d-m_u}},
\end{equation}
where
\begin{equation}
\label{TADPOLE}
\Delta m_K^2\equiv (m_{K^0}^2-m_{K^+}^2)_{QCD} = (6.2 \pm 0.5) 
\times 10^{-3}~\hbox{GeV}^2
\end{equation}
is an estimate for the QCD (or non--photonic) contribution to the 
squared $K^0-K^+$ mass difference. The latter plays a 
central role in both $A(\eta \to \pi^+ \pi^- \pi^0 )$
and the inner bremsstrahlung part of 
$A(\eta \to \pi^+ \pi^- \pi^0 \gamma)$. 
Its numerical value in (\ref{TADPOLE}) is an average between 
the result \cite{GASSER}
$\Delta m_K^2= (m_{K^0}^2-m_{K^+}^2 - m_{\pi^0}^2 + m_{\pi^+}^2) 
= 5.3\times 10^{-3}~\hbox{GeV}^2,$
\noindent
following from Dashen's theorem, and independent estimates 
\cite{BRAMON} (including improved versions of Dashen's theorem 
\cite{DD})  
leading to $\Delta m_K^2$ in the range $(6.5 - 7.0)\times 
10^{-3}~\hbox{GeV}^2$.
\par
The next order lagrangian, $O(p^4)$, contains the Wess--Zumino term 
(the anomalous sector) \cite{WESS} and a series of ten 
(non-anomalous) counterterms identified and studied by 
Gasser and Leutwyler \cite{GASSER}, 
\begin{equation}
\label{L4}
{L}_4= {L}_{WZ} + \sum {L}_i L_4^{(i)}.
\end{equation}
The only pieces of ${L}_{WZ}$ relevant for our purposes are the ones 
containing the anomalous $PPP\gamma$ and $PPPPP$ couplings, i.e.,
\begin{equation}
\label{WZ}
-\sd{e\over 16\pi^2}\epsilon^{\mu\nu\alpha\beta} A_\mu tr ( Q 
\partial_\nu\Sigma\partial_\alpha\Sigma^\dagger 
\partial_\beta\Sigma \Sigma^\dagger -
Q\partial_\nu\Sigma^\dagger\partial_\alpha\Sigma
\partial_\beta\Sigma^\dagger \Sigma) \\
\end{equation} 
and
\begin{equation}
\label{5P}
-\sd{2 \over 15 \pi^2 f^5}\epsilon^{\mu\nu\alpha\beta} 
tr (P \partial_{\mu} P  \partial_{\nu} P  
\partial_{\alpha} P  \partial_{\beta} P),
\end{equation}
respectively.
The finite parts of the ten low--energy constants 
$L_i, i=1,...,10$ are real and have 
been fixed by experimental data \cite{GASSER}. Alternatively, they 
can be approximatively deduced assuming that they are 
saturated by the exchange of known meson resonances 
\cite{REV,BIJNENS1,BIJNENS2,ECKER}, 
thus justifying the phenomenological 
success of conventional VMD. Fixing the renormalization 
mass--scale around these resonance masses ($\mu=M_\rho$, 
for instance), the finite, renormalized values 
for $L_i$ are small enough to justify the convergence 
of the perturbative series. This last remark obviously
does not apply to $L_{WZ}$, generating anomalous processes with 
theoretically well--defined coupling strengths. 

{\bf The inner bremstrahlung amplitude $ A^{(IB)}$}
\par
As previously stated, this part of the amplitude for
$\eta(P) \to \pi^+ (p_+) \pi^- (p_-) \pi^0 (p_0) \gamma(q)$
can be related to the 
purely hadronic one, $A(\eta \to \pi^+ \pi^- \pi^0)$, 
by using Low's theorem. Either through this theorem or by explicit 
calculation, one obtains the following, conveniently factorized 
expression 
\begin{equation}
\label{AIB}
A^{(IB)} = - \frac{B(m_d -m_u)}{3 \sqrt3}  \frac{e}{f^2} 
\left( \frac{\epsilon p_+}{q p_+} - \frac{\epsilon p_-}{q p_-} \right) 
\left( 1 + 2 \frac{m^2_{\eta} - 3Pp_0}{m^2_{\eta} - m^2_{\pi}} 
+ U + V + W \right),
\end{equation}
where the (infrared divergent) factor 
$e \left( \frac{\epsilon p_+}{q p_+} 
- \frac{\epsilon p_-}{q p_-} \right) $, 
containing the photon momentum $q$ and polarization $\epsilon$,
comes from photon radiation by external pions, and the remaining 
factors correspond essentially to  $A(\eta \to \pi^+ \pi^- \pi^0)$
\cite{GASSER}.
\par
The lowest order, $O(p^2)$, contribution proceeds through 
the tree level diagrams shown in Fig. 1a, b and c, with 
vertices of the $L_2$ lagrangian (\ref{L2}).
At this lowest order, all $\eta - \eta'$ mixing effects are ignored and 
the amplitude turns out to be given by eq.(\ref{AIB}) with $U=V=W=0$.
The two diagrams in Fig. 1a give the dominant contribution 
corresponding to the first term in the last parenthesis. 
The second term comes from the three 
diagrams in Fig. 1b with a four-pseudoscalar vertex 
generated exclusively by the derivative part in $L_2$. 
The remaining part of these three diagrams, generated 
by the massive couplings in $L_2$, cancels with the 
contributions from Fig. 1c. However, this lowest order -- 
but otherwise exact -- expression is expected to underestimate the 
$\eta \to \pi^+ \pi^- \pi^0 \gamma$ decay rate since the corresponding 
lowest order amplitude for  $A(\eta \to \pi^+ \pi^- \pi^0)$ 
is known to predict \cite{GASSER} a decay rate well below measurement 
\cite{PDG}.
\par
The next order contribution to $A^{(IB)}$, $O(p^4)$, involves 
one-loop diagrams with vertices from (\ref{L2}) and tree-level 
counter-terms from the non-anomalous part $\sum L_i L_4^{(i)}$
of (\ref{L4}). 
They lead to the $U+V+W$ terms in eq (\ref{AIB}) and their 
values will be taken from the detailed analysis of ref. \cite{GASSER} 
introducing two simplifying approximations. On the one hand, we will 
assume that the smallness of the available phase-space in both 
$\eta \to \pi^+ \pi^- \pi^0$ and $\eta \to \pi^+ \pi^- \pi^0 \gamma$
decays allows to approximate these corrections with the value at the 
center of the Dalitz plot for $\eta \to \pi^+ \pi^- \pi^0$. 
With $\mu = 0.75$ GeV $\simeq M_{\rho}  \simeq M_{\omega} $, this amounts 
to fix $U+V = 0.39-0.03$, coming from pion-loop effects, as discussed 
in detail in \cite{GASSER}. On the other hand, the dominant 
contribution from the finite part of the counter-terms 
$L_i$ is known to come \cite{GASSER} from $\eta -\eta'$ 
mixing effects. In our nonet symmetry context with 
the conventional mixing angle, eq (\ref{MIXING}), this amounts 
to take $1+W \simeq + \sqrt 2$. Notice that this value corresponds to 
a value for $W$ (where $\eta -\eta'$ mixing effects manifest)
somewhat larger than that proposed in \cite{GASSER}, but 
that our simplified treatment is essentially free from the
criticisms recently rised by Leutwyler in ref. \cite{hep}.
With these numerical values and eqs (\ref{B},\ref{TADPOLE}) we obtain 
$\Gamma  (\eta \to \pi^+ \pi^- \pi^0) = 270 \  eV$, in good 
agreement with experiment, 
$\Gamma_{EXP} (\eta \to \pi^+ \pi^- \pi^0) = 283 \pm 27 \ eV$ 
\cite{PDG}. In spite of this agreement, we obviously do not claim that 
our simplified amplitude improves the original and more detailed 
ones in \cite{GASSER} and \cite{hep}. 
We have simply achieved a successful parameterization 
for $\eta \to \pi^+ \pi^- \pi^0$ from which we expect a reasonable 
prediction for $A^{(IB)}$ once inserted in eq (\ref{AIB}).
A similar treatment of the  $\eta \to \pi^+ \pi^- \pi^0$ amplitude can 
be found in a recent analysis by Baur et al. \cite{BAUR}.
\par 
For two different cuts in the photon energy, $E_{\gamma}$, 
we then obtain 
\begin{eqnarray}
\label{WIBs}
\Gamma^{(IB)}(\eta \to \pi^+ \pi^- \pi^0 \gamma) \ =\ 0.050 \ eV&,& \ \ 
 E^{min}_{\gamma} \ =\ 50 \ MeV, \nonumber \\
\Gamma^{(IB)}(\eta \to \pi^+ \pi^- \pi^0 \gamma) \ =\ 0.76 \ \ eV&,& \ \ 
 E^{min}_{\gamma} \ =\ 10 \ MeV,  
\end{eqnarray}
\noindent
and the bremsstrahlung spectrum shown in Fig 2. Uncertainties affecting 
these predictions come mainly from the estimate in eq. (\ref{TADPOLE}) 
and from neglected higher order corrections in ChPT, rather than from 
our simplified treatment of $A(\eta \to \pi^+ \pi^- \pi^0)$. 
Globally, they should be expected to reach some 20 -- 25 \%.
Our results (\ref{WIBs}) are consistent with a previous analysis 
\cite{Fajfer} once the two notations are unified. 

{\bf The Structure Dependent amplitude $A^{(SD)}$}
\par
In contrast with the just discussed bremsstrahlung 
amplitude (\ref{AIB}), which is 
proportional to the isospin violating factor $m_d - m_u$, ChPT predicts 
the existence of further, isospin conserving contributions to the 
global $\eta \to \pi^+ \pi^- \pi^0 \gamma$ amplitude. We now proceed to 
compute the dominant parts of these contributions and collect them into
a structure dependent (or direct emission) amplitude $A^{(SD)}$. 
We will follow the ChPT analysis of $\eta \to \pi^0 \gamma \gamma$ 
\cite{pigg}, whose amplitude
$A(\eta\to \pi^0 \gamma \gamma)$ -- apart from the fact that it 
receives no inner bremsstrahlung contribution -- 
is closely related to $A^{(SD)}$. Indeed, in 
$\eta\to \pi^0 \gamma \gamma$ one photon is isoscalar ($G=-$) and the 
other one is isovector ($G=+$). 
The former is the analog of the  (isoscalar) photon in 
$A^{(SD)}(\eta \to \pi^+ \pi^- \pi^0 \gamma)$, 
while the latter -- having the 
$\rho^0$ quantum numbers -- plays the role of the $\pi^+ \pi^-$ pair. 
Isospin symmetry allows to work in the good 
isospin limit, $m_d=m_u$, and to decompose $A^{(SD)}$ 
in three terms by cyclically rotating the pion charge indeces 
\begin{equation}
A^{(SD)}(\eta \to \pi^+ \pi^- \pi^0 \gamma) \equiv
A^{(SD)} = A^{(SD)}(+,0,-) + A^{(SD)}(0,-,+) + A^{(SD)}(-,+,0).
\end{equation}
\par
The lowest order contribution to $A^{(SD)}$ appears at order four and 
proceeds through the 6 one-loop diagrams of Fig 3, containing two 
vertices from $L_2$, eq (\ref{L2}). Part of this contribution is proportional 
to $m_d - m_u$ and was already included in $A^{(IB)}$, but there is a 
second part -- involving exclusively isospin conserving kaon loops -- 
which belongs to $A^{(SD)}$. This part has to be finite since no 
suitable counterterms are available in (\ref{L4}) for $m_d =m_u$. 
One finds indeed the finite result 
\begin{equation}
A_{(4)}^{(SD)}(+,0,-) = \frac{e}{3 \sqrt6 
\pi^2 f^4} (6Qq + 3Q^2 - 4m^2_K)[(\epsilon p_+)(q p_-) - 
(\epsilon p_-)(q p_+)] I(m^2_K, Q^2, Qq)
\end{equation}
\noindent
with $Q=p_+ + p_-$ and
\begin{equation}
I(m^2_K, Q^2, Qq) \equiv \int_0^1 dx \int_0^{1-x} dy 
\frac{xy}{m_K^2 - 2Qqxy - Q^2 y(1-y)}
\end{equation}
By itself, this order four contribution leads to
\begin {equation}
\label{W4}
\Gamma^{(SD)}_{(4)}(\eta \to \pi^+ \pi^- \pi^0 \gamma) \ 
=\ 0.88 \times 10^{-7} \ eV,
\end{equation}
well below the old estimates (\ref{BR}) and (\ref{WIDTHBR}), and 
suggesting that this is not the dominant contribution to 
$A^{(SD)}$. The same situation was found when analizing 
$O(p^4)$ kaon loops for 
$\eta\to \pi^0 \gamma \gamma$ in ChPT \cite{pigg}. 
Similarly, the smallness of this lowest order contribution seems 
to be a reminiscence of the vanishing of the old CA estimate 
of ref. \cite{Baracca}.
\par
At next order, $O(p^6)$, one has contributions coming from tree level (or counterterms), from 
one loop and from two loops. The former belong to $L_6$ and their 
finite part will be obtained from saturation with vector mesons. 
For further reference, we compute the {\underbar{full}} VMD 
amplitude for $\eta\to \pi^+ \pi^- \pi^0 \gamma$, which proceeds 
through the two diagrams shown in Fig 4 (apart from rotations of pion 
indexes). Ignoring negligible effects from $\Gamma_{\rho, \omega}$
finite widths, one obtains
\begin{eqnarray}
\label{AVMD}
A_{VMD}^{(SD)}(+,0,-) 
= \frac{\sqrt6 e g^4}{4 \pi^4 f^2 (M^2_{\rho} - Q^2)}
\Biggl[ P_{\rho} P_{\omega} \left[ 
 (\epsilon p_+) \left( (P p_-)(qp_0) - (Pp_0)(qp_-) \right) 
\right] + \nonumber   \\  
P_{\rho} P'_{\rho} \left[ (\epsilon p_+) 
\left( (P p_-)(qp_0) - (Pp_0)(qp_-) \right)   
- \frac{1}{3} (qp_0) \left( (\epsilon p_+)(qp_-) - (\epsilon p_-)(qp_+)  
\right) \right] \Biggr] 
\end{eqnarray}
with
\begin{eqnarray}
P_{\rho} P_{\omega} &=&
\frac{1}{M^2_{\omega} - (P-q)^2}
\bigl( \frac{1}{M^2_{\rho} - Q^2} + \frac{1}{M^2_{\rho} - Q^2_+}
+ \frac{1}{M^2_{\rho} - Q^2_-} \bigr) \nonumber \\
P_{\rho} P'_{\rho} &=& \frac{1}{M^2_{\rho} - Q^2} 
\frac{1}{M^2_{\rho} -(P - Q)^2} + \frac{1}{M^2_{\rho} - Q_+^2} 
\frac{1}{M^2_{\rho} -(P - Q_+)^2} + \frac{1}{M^2_{\rho} - Q_-^2} 
\frac{1}{M^2_{\rho} -(P - Q_-)^2} \nonumber 
\end{eqnarray}
and $Q_{\pm} = p_{\pm} + p_0$.
The coupling constants are such that 
$M^2_{\rho} \simeq M^2_{\omega} \simeq 2 f^2 g^2$ and $g=4.2$, 
as discussed in refs.\cite{pigg,BIJNENS1,BIJNENS2}.
The part of this VMD contribution which corresponds to the $L_6$ 
counter-terms in ChPT can be simply obtained from (\ref{AVMD}) by 
expanding the VM propagators
$ 1/ (M^2_V - K^2) = 1/ M^2_V + K^2/ M^4_V $... and retaining only 
the first term. This leads to 
$\Gamma^{(SD)}_{VMD(6)}(\eta \to \pi^+ \pi^- \pi^0 \gamma) \ 
=\ 0.42 \times 10^{-4} \ eV$, which turns out to be 
much larger than the lower order estimate (\ref{W4}).
Since in ChPT loop corrections at order six are expected to be 
only a fraction of the corresponding loop corrections at order 
four, eq (\ref{W4}), we can safely conclude that 
the full ChPT prediction at order six is drastically dominated 
by the VM saturated counter-terms, i.e., 
\begin {equation}
\label{SD6}
\Gamma^{(SD)}_{(6)}(\eta \to \pi^+ \pi^- \pi^0 \gamma) \ \simeq
\Gamma^{SD}_{VMD(6)}(\eta \to \pi^+ \pi^- \pi^0 \gamma) \ 
=\ 0.42 \times 10^{-4} \ eV
\end{equation}
The same dominance of $O(p^6)$ counterterms over the corresponding
loops was also observed 
in the case of $A(\eta\to \pi^0 \gamma \gamma)$ \cite{pigg}.
\par
At order $O(p^8)$ more counterterms appear and a new type of 
loop--correction becomes potentially important. The contributions of 
resonance dominated counterterms can simply be obtained by expanding the
full VMD amplitude (\ref{AVMD}) and retaining only $O(p^8)$ terms.
By itself, this leads to a non negligible order $p^8$ 
correction, namely, 
$\Gamma^{(SD)}_{VMD(8)}(\eta \to \pi^+ \pi^- \pi^0 \gamma) \ 
=\ 0.094 \times 10^{-4} \ eV $
representing a non negligible increase to Eq.(\ref{SD6}).
The new type of loop effects looks {\sl a priori} more 
interesting. Taking two vertices from the anomalous ${L}_{WZ}$
(one from (\ref{WZ}) and another from (\ref{5P})) one obtains
a non--anomalous one--loop correction of order $p^8$, which does not 
vanish only for kaon loops. This "doubly-anomalous" loop 
contributions leads to a decay rate of $ 4.2 \times 10^{-7} \ eV$, 
i.e., of the same order as 
$\Gamma^{(SD)}_{(4)}(\eta \to \pi^+ \pi^- \pi^0 \gamma) \ $ and well 
bellow the order eight counterterm. Again, we find that kaon loops for 
$A^{(SD)}(\eta \to \pi^+ \pi^- \pi^0 \gamma)$ closely follows the same 
pattern as kaon loops for $A(\eta \to \pi^0 \gamma \gamma)$. The whole 
order eight contribution alone is therefore also 
dominated by counterterms,
\begin {equation}
\Gamma^{(SD)}_{(8)}(\eta \to \pi^+ \pi^- \pi^0 \gamma) \ \simeq
\Gamma^{SD}_{VMD(8)}(\eta \to \pi^+ \pi^- \pi^0 \gamma) \ 
=\ 0.094 \times 10^{-4} \ eV
\end{equation}
\par
All this implies that in a ChPT context with resonance saturated
counterterms, the whole structure dependent amplitude for 
$\eta \to \pi^+ \pi^- \pi^0 \gamma $, $A^{(SD)}$, 
is strongly dominated by the 
contributions of these counterterms and should essentially be given by 
the full VMD amplitude (\ref{AVMD}). Such an "all-order" estimate leads 
to 
\begin {equation}
\label{VMDW}
\Gamma^{SD}(\eta \to \pi^+ \pi^- \pi^0 \gamma) 
\simeq 
\Gamma_{VMD}(\eta \to \pi^+ \pi^- \pi^0 \gamma)  
=\ 1.4 \times 10^{-4} \ eV
\end{equation}
The corresponding photonic spectrum is shown in Fig. 2.
\par
This structure dependent part of the amplitude is the one that should 
be compared to older estimates, although none of those treatments 
coincides precisely with ours. From eq (\ref{VMDW}) and the result
$\Gamma_{VMD}(\eta \to \pi^0 \gamma \gamma)  =\  0.31 \ eV$ 
of ref. \cite{pigg}, one obtains 
$\Gamma_{VMD}(\eta \to \pi^+ \pi^- \pi^0 \gamma) /
\Gamma_{VMD}(\eta \to \pi^0 \gamma \gamma) \simeq 0.44 \times 10^{-3} $
somewhat below the old estimates (\ref{BR}) and above the vanishing 
prediction of ref. \cite{Baracca}.
The shape of our photonic spectrum coincides with the old
prediction by Singer \cite{SINGER} who used a simplified version of 
VMD not far from ours.
\par
Our final ChPT predictions for the {\it whole}
$\eta \to \pi^+ \pi^- \pi^0 \gamma$ amplitude can be obtained 
from the sum of $A^{(IB)}$, eq (\ref{AIB}), and the full VMD amplitude,
eq (\ref{AVMD}). 
The contribution of the latter -- in modulus plus interference with 
$A^{(IB)}$ -- has negligible effects for soft photons and 
represents a minor increase (somewhat below 1 \%)
of the decay rate in the higher half of the photonic spectrum as shown 
in Fig. 2. The integrated width remains thus unaffected as in 
eq (\ref{WIBs}).
\par
{\bf Conclusions}
\par
From the preceding and somewhat intricate analysis of the 
$\eta \to \pi^+ \pi^- \pi^0 \gamma$ decay, a definite two-fold 
conclusion emerges, namely, that the whole amplitude is 
strongly dominated by inner bremsstrahlung and that it is expected to 
be large enough to allow for detection in a near future. 
This is a useful conclusion in that it contradicts (and presumably 
corrects) older results (see refs. \cite{INTEMANN},\cite{SINGER}) and 
it will furnish a clear-cut test for (and presumably confirm) ChPT. 
But it is also a deceptive conclusion reducing most of the
$\eta \to \pi^+ \pi^- \pi^0 \gamma$ dynamics to that of 
$\eta \to \pi^+ \pi^- \pi^0$, already studied both theoretically 
\cite{GASSER} and experimentally \cite{PDG} (although not free of 
uncertainties \cite{GASSER,hep,DONOGHUE}). 
Only an extremely sensitive and dedicated experiment, and an 
improved theoretical treatment of $A^{(IB)}$ could 
allow to extract the structure dependent part of the amplitude which 
contains the genuinely new effects of ChPT for 
$\eta \to \pi^+ \pi^- \pi^0 \gamma$. 
In this case, the values of counterterms  and the hypothesis of their
resonance saturation -- rather than chiral loop effects -- will be 
tested. Stated otherwise, the experimental analysis of 
$\eta \to \pi^+ \pi^- \pi^0 \gamma$ can represent an excellent 
confirmation of ChPT if the predicted spectrum (largely dominated by 
bremsstrahlung) is observed, but it can hardly be useful to improve our 
knowledge on other aspects of this theory .

\par
{\bf ACKNOWLEDGEMENTS}
\par
Thanks are due to Ll. Ametller, E. Bagan, F. Cornet and S. Peris 
for clarifying discussions.  P.G. acknowledges gratefully grants from
the Generalitat de Catalunya and  the Ministerio de Educaci\'on y
Ciencia. 
This work has partially been supported by CICYT, AEN95-0815, and by 
EURODAPHNE, HCMP, EEC contract \#CHRX-CT920026.


\newpage

\begin{figure}
\centerline{
            \epsfig{file=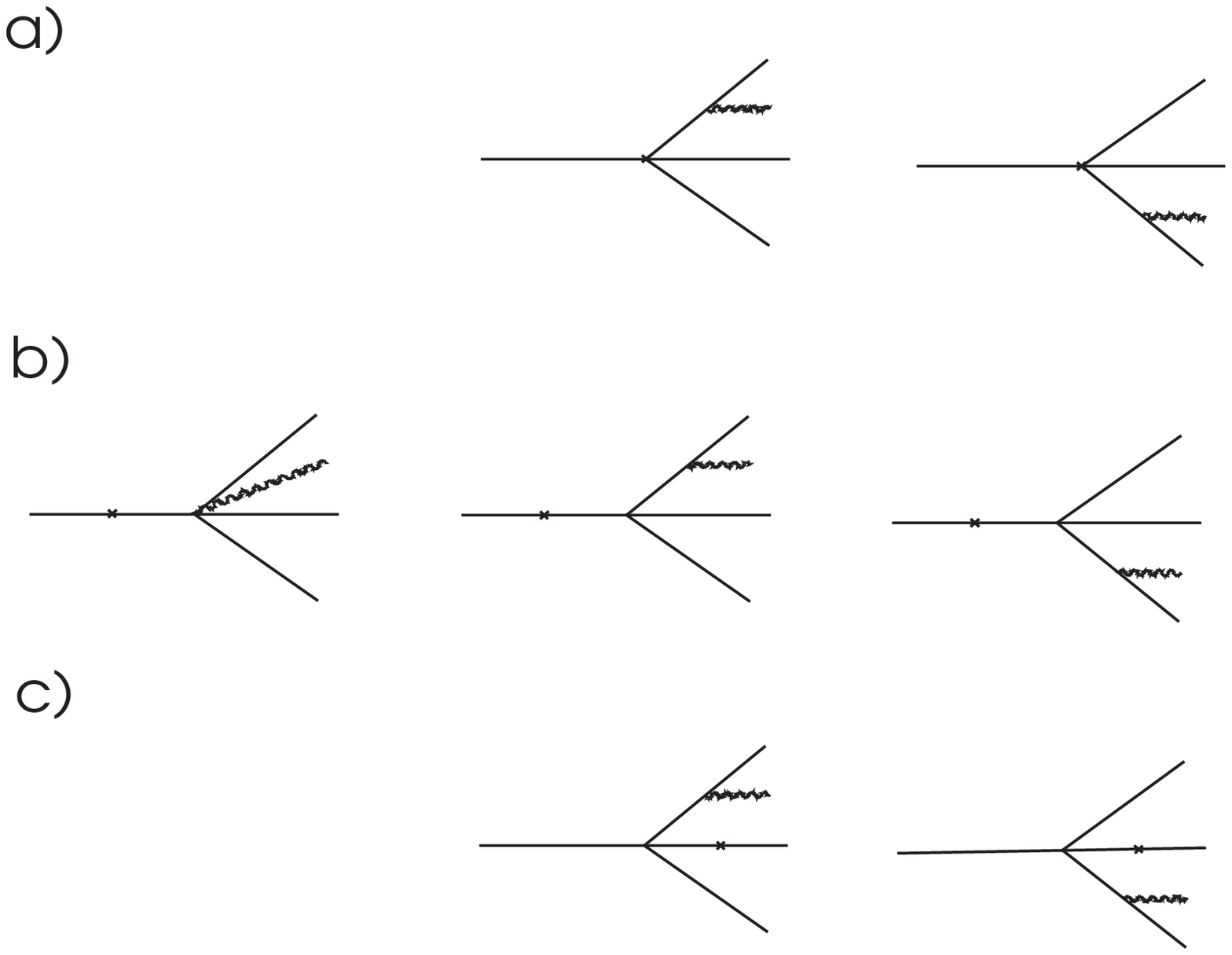, height=10cm, width=10cm}
           }
            \caption{
Diagrams contributing to $A^{IB}(\eta\to \pi^+\pi^-\pi^0\gamma)$ at 
lowest order, $O(p^2)$, in ChPT. The cross in each diagram indicates a 
coupling proportional to $m_d - m_u$.
\label{cacatua1} }
\end{figure}

\begin{figure}
\begin{tabular}{rl}
\mbox{${d \Gamma \over d E_\gamma}$} &
\mbox{ $\vcenter{
\epsfig{file=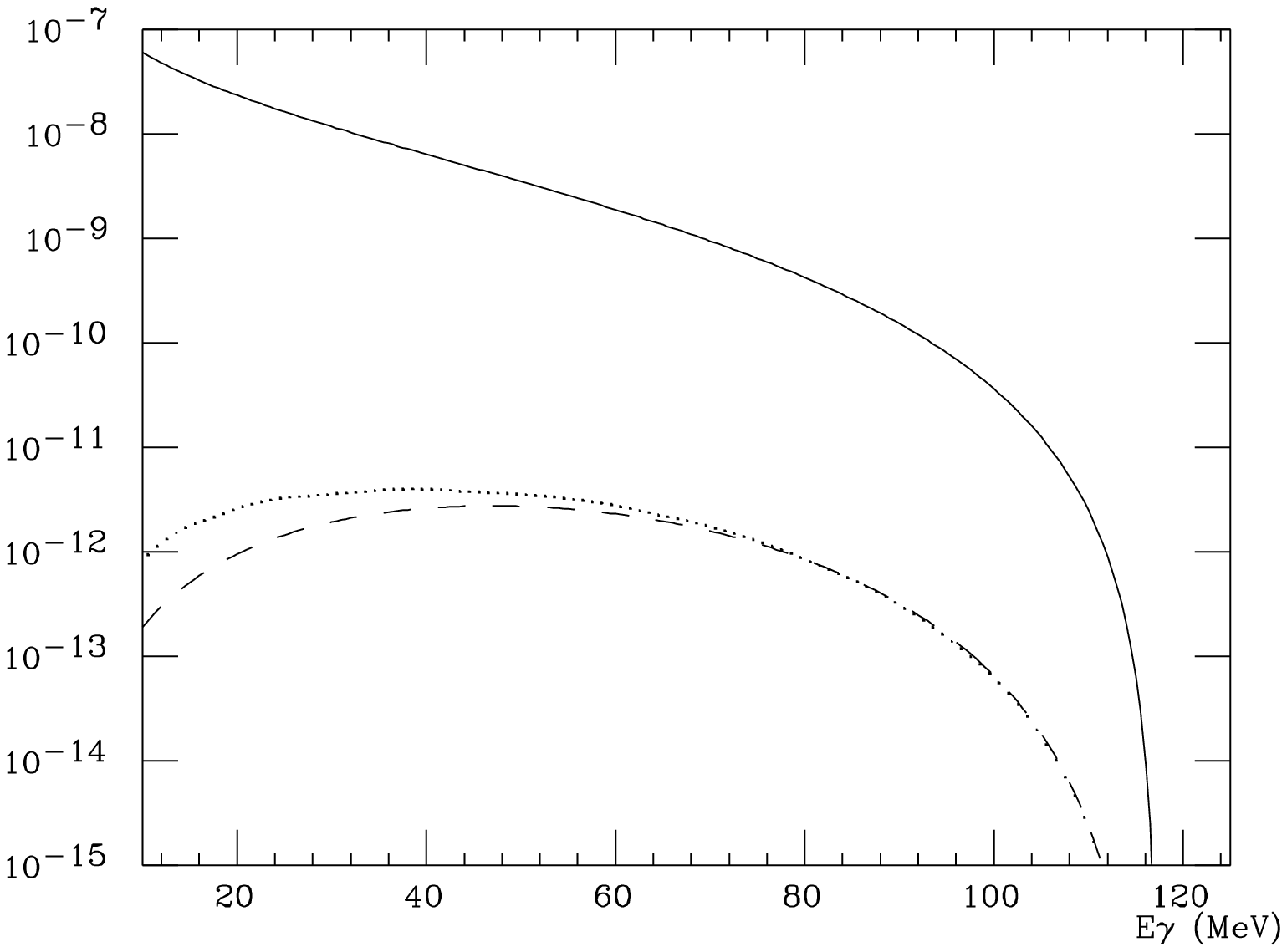, height=10cm, width=10cm}
}$
}
\end{tabular}
            \caption{
Spectrum of the photon energies, $E_\gamma$, as predicted by 
ChPT for the $\eta\to \pi^+\pi^-\pi^0\gamma$ decay. 
The solid line corresponds to the inner bremsstrahlung contribution.
The dashed line is the VMD contribution alone. 
The dotted line is their interference. 
\label{cacatua2} }
\end{figure}

\begin{figure}
\centerline{
            \epsfig{file=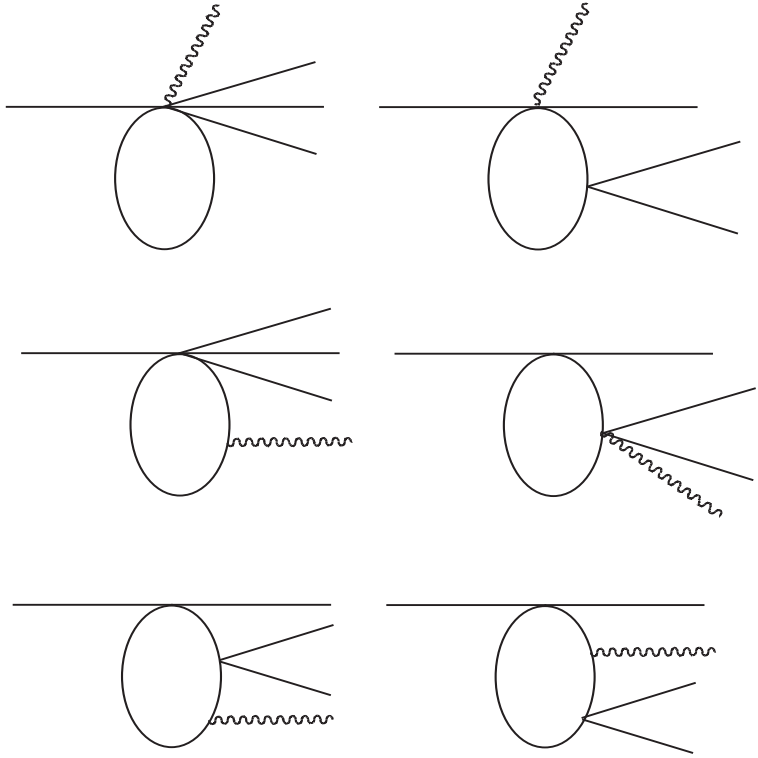, height=10cm, width=10cm}
           }
\caption{
Kaon loops contributing to $A^{SD}(\eta\to \pi^+\pi^-\pi^0\gamma)$
at $O(p^4)$. 
\label{cacatua3} }
\end{figure}

\begin{figure}
\centerline{
            \epsfig{file=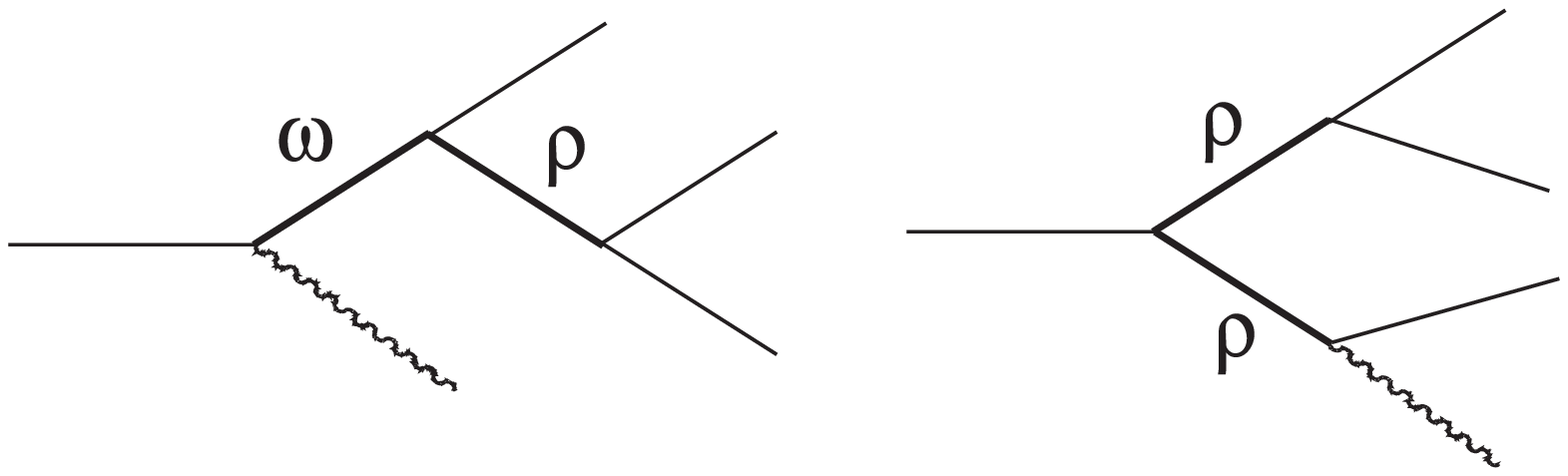, height=4cm, width=10cm}
           }
\caption{
VMD contributions (apart from pion index rotations) to 
$A^{SD}(\eta\to \pi^+\pi^-\pi^0\gamma)$.
\label{cacatua4} }
\end{figure}


\newpage


\begin{thebibliography}{99}
\bibitem{GASSER}
J. Gasser and H. Leutwyler, Nucl. Phys. {\bf B250} (1985) 465 and 539.
\bibitem{DASHEN}
R. Dashen, Phys. Rev {\bf 185} (1969) 1248. \\
R. Dashen and M. Weinstein, Phys. Rev {\bf 185} (1969) 1291.
\bibitem{SINGER}
P. Singer, Phys. Rev. {\bf 154} (1967) 1592.
\bibitem{INTEMANN}
G. Intemann and I. Lapidus, Phys. Rev. {\bf 165} (1967) 1650; \\
\bibitem{old}
A. Q. Sarker, Phys. Rev. Lett. {\bf 19} (1967) 1261; \\
\bibitem{Baracca}
A. Baracca et al., Nuovo Cimento,{\bf 50} (1976) 1006.
\bibitem{PDG}
Particle Data Group, Phys. Rev. {\bf D50} (1994) 1173.
\bibitem{THALER}
J. J. Thaler et al., Phys. Rev. {\bf D7} (1973) 2569.
\bibitem{pigg}
Ll. Ametller et al., Phys. Lett. {\bf B276} (1992) 185.
\bibitem{REV}
J. Bijnens, Int. J. Mod. Phys. {\bf A8} (1993) 3045. \\
Ll. Ametller, The Second Daphne Physics Handbook, ed. by 
L. Maiani, G. Pancheri and N. Paver, INFN, Frascati 1995.
\bibitem{DONOGHUE}
J. F. Donoghue, B. R. Holstein, Y.C.R. Liu, Phys. Rev. Lett. {\bf 55} 
(1985) 2766. \\
J. Bijnens, A. Bramon and F. Cornet, Phys. Rev. Lett. {\bf 61} 
(1988) 1453.
\bibitem{BIJNENS1}
J. Bijnens, A. Bramon and F. Cornet, Z. fur Phys. {\bf C46} 
(1990) 595. 
\bibitem{BIJNENS2}
J. Bijnens, A. Bramon and F. Cornet, Phys. Lett. {\bf B237} 
(1990) 488.
\bibitem{BRAMON}
A. Bramon and E. Mass\'o, Phys. Lett. {\bf B93} (1980) 65.
\bibitem{DD} 
J. F. Donoghue et al., Phys. Rev. Lett. {\bf 69} (1992) 3444; \\
J. Bijnens, Phys. Lett. {\bf B306} (1993) 343.
\bibitem{WESS}
J. Wess and B. Zumino, Phys. Lett. {\bf 37B} (1971) 95. \\
E. Witten, Nucl Phys. {\bf B233} (1983) 422.
\bibitem{ECKER}
G. Ecker, J. Gasser, A. Pich and E. de Rafael, Nucl. Phys. {\bf B321} 
(1989) 311.
\bibitem{hep}
H. Leutwyler, BUTP-96/5 preprint and hep-ph-9601236.
\bibitem{BAUR}
R. Baur, J. Kambor and D. Wyler, ZU-TH 16/95 and IPNO/TH 95-53 
preprints.
\bibitem{Fajfer}
S. Fajfer et al., Phys. Rev. {\bf D44} (1991) 295.
\bibitem{DONOGHUE}
J. F. Donoghue, E. Golowich and B. Holstein, "Dynamics of the Standard 
Model", CUP, Cambridge 1992.


\end{thebibliography}
\end{document}